\documentclass[runningheads]{llncs}
\usepackage{comment}
\usepackage{appendix}
\usepackage{subcaption}
\usepackage[]{graphicx}

\usepackage{float}

\usepackage{placeins}
\usepackage{hyperref}
\usepackage[inline]{enumitem}
\usepackage{algorithm}

\begin{document}

\title{Temporal Analysis of Worldwide War}
%
%\titlerunning{Abbreviated paper title}
% If the paper title is too long for the running head, you can set
% an abbreviated paper title here
%
\author{Devansh Bajpai \and Rishi Ranjan Singh}
\authorrunning{Bajpai and Singh}
% First names are abbreviated in the running head.
% If there are more than two authors, 'et al.' is used.
%
\institute{Department of Electrical Engineering and Computer Science\\
Indian Institute of Technology, Bhilai\\
Chhattisgarh, India. \newline
\email{\{devanshb,rishi\}@iitbhilai.ac.in}}

\maketitle     
%%%%%%%%%%%%%%%%%%%%%%%%%%%%%%%%%

\begin{abstract}
 Analysis of wars and conflicts between regions has been an important topic of interest throughout the history of humankind. In the latter part of the $20^{th}$ century, in the aftermath of two World Wars and the shadow of nuclear, biological, and chemical holocaust, more was written on the subject than ever before. Wars have a negative impact on a country's economy, social order, infrastructure, and public health. In this paper, we study the wars fought in history and draw conclusions from that. We explore the participation of countries in wars and the nature of relationships between various countries during different timelines. A big part of today’s wars is fought against terrorism. Therefore, this study also attempts to shed light on different countries’ exposure to terrorist encounters and analyses the impact of wars on a country's economy in terms of change in GDP.
\end{abstract}

%%
%% The code below is generated by the tool at http://dl.acm.org/ccs.cfm.
%% Please copy and paste the code instead of the example below.
%%

%% Keywords. The author(s) should pick words that accurately describe
%% the work being presented. Separate the keywords with commas.

\section{Introduction}
Wars are one of the most important factors in deciding the state of the world. These are the major turning points in any country or empire’s history. Several literature study about cause~\cite{Organski:1981,Henk:1988,Mcmahan:2005}, damage~\cite{Murthy:2006,Glick:2010} and outcomes~\cite{David:1999,Karl:2004} associated with interstate and civil wars. In the network of military alliances, wars and international trade, the relation between the international trade and wars happening among the countries have been studied~\cite{Hafner:2009,Jackson:2015}.  In \cite{Lagazio:2004}, authors have done a neural network based analysis of militarized disputes for a timeline from 1885 till 1992. Network theoretic analysis for international relations is done in \cite{Hafner:2009}. A similar work to this paper \cite{Homero:2018}, analysed  temporal network of international relations based on the wars fought between 1816 and 2007. Since the events are historical we already know the facts. Therefore, it becomes easier to connect the reported results with the cause that triggered it.

In this paper, we carry out a temporal analysis of the wars that have been fought in the history and report the inter-country relationships during different timelines. We focus only on the wars fought after 1500 CE due to availability of data and with the assumption that wars fought before 1500 CE might have only minor effects on the present inter-country relationships. We use temporal multi-graph to analyze the wars fought during 1500-2020 timeline. %For phase-wise analysis, further the timeline 1500-2020 is divided into 3 timelines: \textit{early wars [1500-1800], during and pre-world wars[1801-1945],} and \textit{post-world wars [1946-2020]}. 
Some basic graph properties are used for the analysis of temporal data. The distribution of wars over the timeline, countries participating in most of the wars in different timelines, rival relationship between countries, countries which fought most wars along the same side during different timelines are few of the focal points of this study. The relevant statistics and crucial conclusions are reported. Some of the major events related to wars in the history are marked and also mapped to respective roots and causes. %The inter-country relation is evaluated by studying the number of wars between them, number of war fought against same enemy and number of wars fought together by them. 
Terrorist organizations are also considered as nodes in the temporal graphs to analyze the terrorist activities and encounters with different countries. %Details about most active terrorist organizations, and countries which have been most exposed to terrorist encounters, is also analyzed. 
Finally, the impact of wars on a country's economy has been studied in terms of change in GDP.

\section{Data Collection}
One of the main reasons for limited related research is the limited availability of data-sets. A data-set maintained by Sarkees and Wayman\cite{Sarkees:2010} covers the wars from  1816 - 2007 timeline. It only covers a small portion of time-line that we plan to cover. Their study also did not cover the impact of wars on economy of countries. Therefore, we create a data-set for more wider timeline which we collect based on the data available on Wikipedia pages~\cite{wiki}. These set of pages list war history of 165 different countries. Each country has a different page listing wars and conflicts that the country has participated in. The temporal multi-graph was built based on the data available on these pages. \textit{Requests} and \textit{Beautiful Soup} libraries available with Python3 are used for scraping the web pages. These libraries can fetch the content written inside HTML tags. A python script is used which processes every country’s page and scrapes the details of the wars that country has participated in. Most of the pages have the same format and contain information such as, name of wars, countries participated in wars and timelines of wars. The raw data from most of the pages is retrieved and stored in a CSV file. Next, the raw data collected in CSV file is used to create a temporal multi-graph from it.
 
 The MPD~\cite{https://doi.org/10.1111/1468-0289.12032,10.2307/27744128,RePEc:cup:cbooks:9780521817912,BROADBERRY201558,RePEc:hit:hitcei:2018-13,https://doi.org/10.1111/1475-4932.12250,pfister2011economic,malanima2011long,baffigi2011italian,broadberry2010british} is a
 freely available excel spread-sheet (Maddison, 2016) that provides per capita income for 184 countries upto 2016 CE. We have used the MPD data to study the impact of wars on country's economic growth. The data for years 1850-2016 has been used since most of the values were missing for years before 1850.

\subsection{Challenges in Network Formation}
Out of 165 around 25 pages had a different format. The difference was in the formatting of available data. Since the scraping script is automated for general format, therefore, it failed to process those 25 pages. After some further modifications in the script and manually writing some entries directly in CSV file, the complete data was collected.  

Entries in the CSV file with different formats made it difficult to directly read data from the CSV file and create the network. For example, different formats used for representing timeline shows why temporal network could not be made directly: 1948, 1948-1948, May 1948 September 1948, 5th May 1948-9th September 1948. 18th century, Early 17th century to 19th century.\\
A data processing script is used to process each input and convert it into a standard format. Python’s regular expression library \textit{RE} and some string processing are used in this script. Next, an integer id is assigned to each opponent, as well as a node is added for it and the adjacency list is created. 

Some countries are known by many names and a separate node was created for all of them. For example, Argentina was mentioned as: Argentine Republic, Argentine Armed Forces, Argentine Air Force, Exiled Argentines, Argentine Army, Argentine Navy, Argentine Unitarians, Government of Argentina and  Argentine Confederation i.e. 9 different names. To counter this issue, another processing layer was used before assigning an id to an opponent as well as adding a node for it, which replaced each opponent's name from a map entry. Map contained standard names for countries which were present with multiple names in our data-set. The map was created manually after observing entries in the data-set.

\section{Network Description}
Nodes in the network represent the fighting entity which could be a nation(e.g. India), an empire (e.g. Ottoman empire), countries or empires which existed in history (e.g. Saxony), terrorist organisations (e.g. ISIS), inter-governmental country alliances (e.g. NATO). Edges of the network represent the wars fought between nodes. Network is in the form of a temporal multi-graph with a total of 3000 nodes and 27721 edges. The average degree of the graph, considering edges over the whole timeline, is 18.5. The maximum degree is 1321 and the minimum degree is 1. Here, the degree of a node represents the number of wars that country/empire/terrorist organization was a part of.

\begin{figure}[htbp]
  \centering
  \includegraphics[width=.5\linewidth]{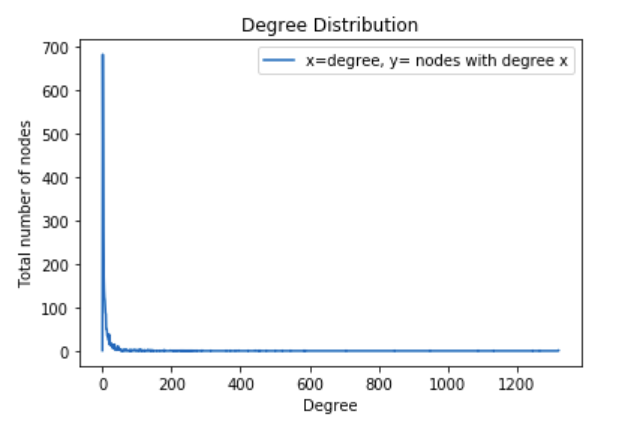}
  \caption{Degree Distribution of The Network}
%Description{The 1907 Franklin Model D roadster.}
  \label{deg_dist}
\end{figure}

In Figure~\ref{deg_dist}, the degree distribution has been plotted, where the $x$-axis represent the degree of a node, i.e., the number of wars a country/empire/terrorist organization was a part of. The $y$-axis denotes the number of nodes with degree $x$, i.e., the number of countries/empires/ terrorist organizations with $x$ number of war participation. 

\begin{table}[htbp]
  \caption{Degree Distribution}
  \label{tab:freq}
  \begin{center}
 {\textbf{\footnotesize \begin{tabular}{|c|c|l|}
    \hline
    Total Nodes &3000\\
   \hline
    Number of nodes with degree 1 & 682\\
    Number of nodes with degree $\le$ 10& 2205\\
    Number of nodes with degree $\le$ 50& 2832\\
   Number of nodes with degree $\le$ 100& 2896\\
    Number of nodes with degree $\le$ 300& 2975\\
    Number of nodes with degree $\ge$ 301& 25\\
  \hline
\end{tabular}}}
    
\end{center}
\label{tab_deg}
\end{table}
 Degree distribution of the network in the presence of all edges is skewed. The degree distribution curve follows power law. Table~\ref{tab_deg} summarises the cumulative degree for some of degree values. It is clear from the table that only $0.833\%$ of the total number of node have more than 300. $73.5\%$ of the nodes are with at most 10 degree. Very few entities have been there that have been engaging in a lot many number of wars. A large number of countries/empires/ terrorist organizations do not participate in several wars.In next section, we identify these nodes with high degree over different timelines. 

%%Graph has two versions, weighted and unweighted. I have used different weight metrics with general format:\\
%%start year = starting year of that war.\\
%%end year = ending year of that war.\\
%%para= parameter value taken from (100,1000,end year -start year)\\

%%weight= start year + para*(end year-start year)\\
%%And finally for normalising the values each weight was divided by 1500.\\

 %In Figure~\ref{deg_dist}, the degree distribution has been plotted, where the $x$-axis represent the degree of a node, i.e., the number of wars a country/empire/terrorist organization was a part of. The $y$-axis denotes the number of nodes with degree $x$, i.e., the number of countries/empires/ terrorist organizations with $x$ number of war participation. A large number of countries/empires/ terrorist organizations do not participate in several wars.

%It is clear from Figure~\ref{deg_dist} that the plotted curve follows power law distribution. Only $0.833\%$ of the total number of node have degree more than 300. $73.5\%$ of the nodes are with at most 10 degree. 

\section{Results and Discussion}
In this section, we note down all the statistical findings as well as conclusions drawn from these over the considered timeline. 
\subsection{Timeline analysis:} 
\begin{figure}[tbp]
\begin{subfigure}{0.325\textwidth}
\includegraphics[width=\textwidth]{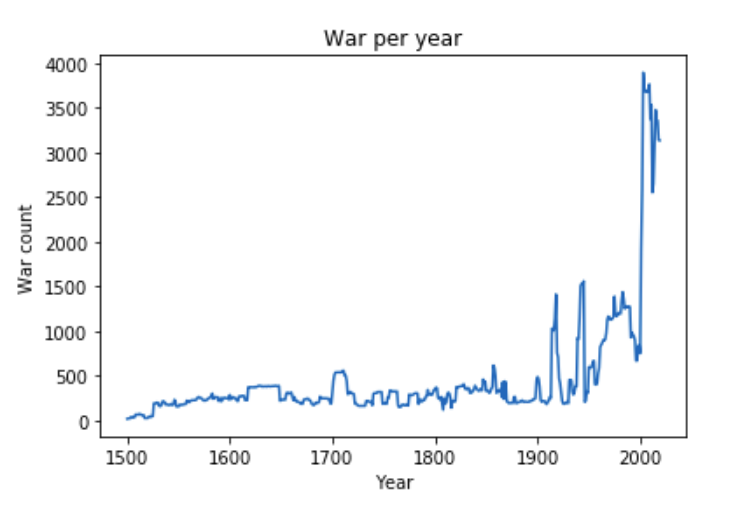}
\caption{Considering all nodes}
\label{fig_year}
\end{subfigure}%
\begin{subfigure}{0.325\textwidth}
\includegraphics[width=\textwidth]{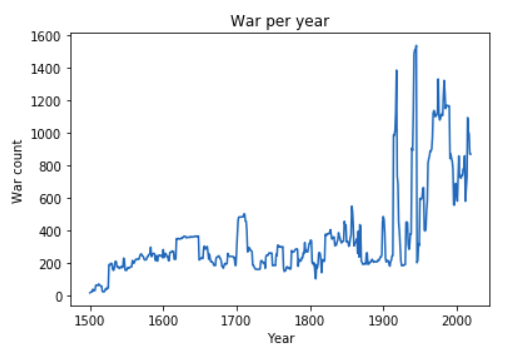}
\caption{Ignoring terrorist units}
\label{fig_year_wto}
\end{subfigure}
\begin{subfigure}{0.325\textwidth}
\includegraphics[width=\textwidth]{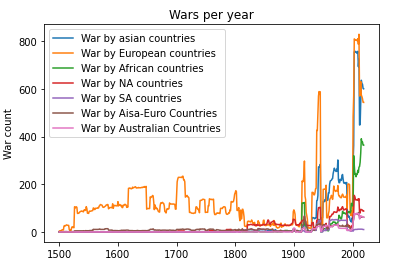}\\
\caption{For each continent}
\label{fig_cont}
\end{subfigure}
\caption{Year-wise number of wars}
\end{figure}

We have considered wars that have happened between 1500 CE and 2020 CE. It is a period of 520 years. Analysis of the whole timeline at once doesn't give us good results as an ally in the fifteenth century might be enemy in the twenty first century. It is also difficult to do analysis on each year separately as there are a total of 520 years. To decide the segments, we created a plot Figure~\ref{fig_year} between years and number of wars fought in a specific year. In Figure~\ref{fig_year}, It can be observed there is a huge increase in the number of wars after 2001. In particular, the height of the curve at year 2003 is 3895 which is more than twice of the cumulative number of wars during the second world war. We found that many countries had around 30 edges every year because of encounters with different terrorist organisations like Tahrik-i-Taliban Pakistan, HI-Khalis, Islamic Emirates, Boko Haram etc. We drew, another plot, given in Figure~\ref{fig_year_wto}, similar to Figure~\ref{fig_year}, but without the terrorist encounters. In the new plot total wars fought reduced heavily, e.g. wars  fought in 2016, 2017, 2018 and 2019 reduced from 3351, 3370, 3153, 3148 to 1008, 994, 871 and 870 respectively. Which are less than 2/3 of the wars fought during the second world war.  In Figure~\ref{fig_year_wto}, it is observed that the number of wars are almost constant before 1800. After, 1800 spikes start occurring in the curve. There are 3 big spikes after 1800 and before 2000. First spike is at year 1914 which is in the timeline of the first world war, and there are 1411 edges in that year. Second spike is at year 1945 which is in timeline of the second world war, and there are 1560 edges in that year. Third spike is at year 1987. After analyzing the edges, it was found out that this spike is because of the different individual wars which happened in 1987 e.g. Iran-Iraq war, Cold War, Baltic states vs Soviet Union etc. %Therefore, we divided this timeline from 1500 CE to 2020 CE into three smaller timeline segments. 

After analyzing the plot in Figure~\ref{fig_year} and Figure~\ref{fig_year_wto}, we concluded to divide the timeline from 1500 CE to 2020 CE into three parts and study all of them separately as well.
 
\begin{enumerate}
   \item Early wars[1500-1800]: This is the part in the graph which has almost same number of wars. The wars are mostly dominated by European countries. 4413 wars started in this timeline, which is least of the three timelines.
    \item During and pre-world wars[1801-1945]: This timeline covers the incidents of World wars. Both the world wars fall in this timeline, 9064 wars started in this timeline and total 9510 war were fought in this timeline.
    \item Post world wars[1946-2020]: Post world war era. This timeline's results are most effective in influencing the present state of different countries. 14244 wars started in this timeline and 14358 wars were fought in this timeline.
\end{enumerate}
\textbf{Wars in different Continents:} We also plot the number of wars fought by countries from different continents. The plot is given in Figure~\ref{fig_cont} where the $x$-axis denotes the years and the $y$-axis denotes the number of wars fought by countries from a continent. For each continent, a curve with a different color is plotted. The nodes having degree more than 40 are only considered for respective continents. We have only plotted curve for Asia, Europe, Africa, North America, South America and  Australia. We haven't included Antarctica as none of the country with degree more than 40 belonged to it. We also added a continent with name Euro-Asia to cover the countries which exist partially in Asia and partially in Europe. From Figure~\ref{fig_cont}, we can observe that European countries dominated the world in terms of participating in wars between 1500 to 1800. 1800-1900 period had relatively lesser number of war. This period wasn't solely dominated by European countries as other continents' degree had increased. Number of wars increased significantly during the world war timelines and European nations had the highest degree. Asia, Europe and Africa continents have mostly participated in wars after 2000.

\subsection{High Degree Nodes}
The nodes having high degrees in the constructed network correspond to the countries or empires fighting most number of wars in that timeline. The powerful countries generally become condescending and fight more wars. So, the high degree nodes in the graph represent the powerful countries of the timeline. Table~\ref{tab_t_deg} shows the list of top 10 countries/empires that participated in maximum number of wars during different timelines.\\ 

\begin{table}[bp]
\caption{High Degree Nodes}
\resizebox{\textwidth}{!}{
\begin{tabular}{|lc|lc|lr|lc|}
\hline
\multicolumn{ 2}{|c|}{\textbf{[1500 – 1800]}} & \multicolumn{ 2}{c|}{\textbf{[1801 – 1945]}} & \multicolumn{ 2}{c|}{\textbf{[1946 – 2020] }} & \multicolumn{ 2}{c|}{\textbf{[1500 – 2020] }} \\ \hline
\multicolumn{1}{|c|}{\textbf{Node}} & \textbf{Degree} & \multicolumn{1}{c|}{\textbf{Node}} & \textbf{Degree} & \multicolumn{1}{c|}{\textbf{Node}} & \multicolumn{1}{c|}{\textbf{Degree}} & \multicolumn{1}{c|}{\textbf{Node}} & \textbf{Degree} \\ \hline
\hline
Ottoman Empire & \textbf{376} & Russia & \textbf{608} & United States & \textbf{632} & Russia & \textbf{1279} \\ \hline
France & \textbf{364} & Germany & \textbf{587} & Russia & \textbf{455} & United Kindom & \textbf{1276} \\ \hline
United Kindom & \textbf{334} & Italy & \textbf{564} & United Kindom & \textbf{433} & France & \textbf{1253} \\ \hline
Spain & \textbf{333} & France & \textbf{543} & Al-Qaeda & \textbf{410} & United States & \textbf{1065} \\ \hline
Russia & \textbf{237} & United Kindom & \textbf{526} & France & \textbf{351} & Germany & \textbf{999} \\ \hline
Holy Roman Empire & \textbf{207} & United States & \textbf{424} & Australia & \textbf{334} & Spain & \textbf{865} \\ \hline
Turkey & \textbf{199} & Ottoman Empire & \textbf{405} & Germany & \textbf{292} & Italy & \textbf{848} \\ \hline
Dutch Republic & \textbf{191} & Japan & \textbf{330} & China & \textbf{287} & Ottoman Empire & \textbf{781} \\ \hline
Sweden & \textbf{185} & Spain & \textbf{324} & Iraq & \textbf{285} & Romania & \textbf{583} \\ \hline
Portugal & \textbf{161} & Bulgaria & \textbf{316} & Italy & \textbf{250} & Bulgaria & \textbf{495} \\ \hline
\end{tabular}}
\label{tab_t_deg} 

\end{table}

\noindent \textbf{Timeline 1: Early Wars[1500-1800]:} This was an era where countries had to fight against each other to establish their control and show dominance. We can observe in the [1500-1800] section of Table~\ref{tab_t_deg} Ottoman Empire is the node with highest degree. During 16th and 17th centuries, the Ottoman Empire was a multinational, multilingual empire controlling most of Southeast Europe, parts of Central Europe, Western Asia, parts of Eastern Europe and the Caucasus and North Africa. The presence of its name at the top of the list reflects its dominance in the timeline. Most of the nodes in the list are European countries as they were participating in a lot of wars in trying to establish their colonies.\\

\noindent \textbf{Timeline 2: Pre-world wars [1801-1945]:} In the first world war, Germany, Ottoman Empire and Bulgaria were the major participants of the first side which were opposed  by Russia, United Kingdom. In the second world war, Germany, Italy and Japan were major participants of the first side which were opposed by France, United Kingdom, Russia and United States. Clearly all the nodes except Spain in the [1801-1945] section of Table~\ref{tab_t_deg} are among the major participants of one of the world war. Russia, Germany, Italy, France and United Kingdom actively participated in both the world wars so they are at the top of list, where as United States, Ottoman Empire and Bulgaria were active in one of the world wars so they are towards the bottom of list. Spain is present in the list as it also participated in WW2 and also fought wars during 1800-1900. \\

\noindent \textbf{Timeline 3: Post-World wars[1946-2020]:} United States is at the top of the [1946-2020] section of Table~\ref{tab_t_deg} as it is known for its efforts in trying to mitigate terrorism, US army is fighting against terrorism in various areas like Afghanistan, Iraq and Syria. USA also fought the cold war against Russia. It is followed by other countries like Russia, United Kingdom, France, Australia etc, all of these nations are undoubtedly powerful nations of the present era. This is the only timeline in which we see a terrorist organisation (al-Qaeda) present in the list, which supports the fact that this timeline is dominated by the wars against terrorism.\\

\noindent \textbf{Overall [1500-2020]:} %The nodes having high degrees in our graph correspond to the countries or empires fighting most number of wars in last 520 years. Earlier the nodes which were present at the top of high degree node lists were the nodes which dominated the timeline. 
Now, since the timeline is much bigger the nodes present in the list could be the node which were powerful throughout the timeline or the nodes which dominated a specific timeline. Nodes like Russia, United Kingdom, France, Germany are the ones which were powerful in all three timelines as we have seen in the timeline specific analysis. Whereas nodes like United States, Ottoman Empire and Bulgaria were not very strong in all three timelines. For example, USA was not active in the first timeline but dominated the next two, Ottoman Empire on the other side was powerful in the first two timelines and disappeared from the list in the third timeline. 
\subsection{Rival Countries}
The nodes having highest number of edges between them in our graph corresponds to the pair of countries fighting most number of wars against each other during a timeline. List of pair of nodes with highest number of edges between them gives us the list of countries with strong rivalries between them. Table~\ref{tab_riv_t} shows the list of such rival countries.\\

\noindent \textbf{Timeline 1: Early Wars[1500-1800]:}  We can observe in [1500-1800] section of Table~\ref{tab_riv_t} that all of these rivalries are among the European Countries. United Kingdom-France is at the top of the list. Anglo-French wars which were happening for almost throughout this timeline are example of conflicts between England and France. Similarly all other pairs are also of European countries fighting for dominance.\\

\noindent \textbf{Timeline 2: Pre-world wars [1801-1945]:} Germany, Ottoman Empire and Bulgaria fought against Russia, United Kingdom in world war 1. Germany, Italy and Japan fought against France, United Kingdom, Russia and United States. United kingdom and Russia fought against Germany in both the world wars rooting for the first two entries of [1801-1945] section of Table~\ref{tab_riv_t}. United kingdom and Russia fought against Italy in the second world war rooting for the next two entries of table. Indonesian national revolution was a major armed conflict that happened in 1945-1949 which included a number of military encounters between Dutch and Indonesia rooting for the presence of last entry in the list. \\

\noindent \textbf{Timeline 3: Post-World wars[1946-2020]:} As mentioned earlier this timeline is the most recent and stats of this timeline are most effective in deciding the current situations. All the rival nodes present in the [1946-2020] section of Table~\ref{tab_riv_t} are well known rivals of present era. United States and Russia fought long cold war against each other. United states-china, Thailand-Vietnam, China-Vietnam our well known conflicts. Al-Qaeda attacked on United States in 2001 and straight after that USA declared war against it which is still going on.\\

\noindent \textbf{Overall [1500-2020]:} The [1500-2020] section of Table~\ref{tab_riv_t} shows the pairs of nodes which constantly fought against each other or the pairs of nodes which fought many battles against each other in specific timeline. We can see that United Kingdom- France is at the top of the list because they fought against each other in the first timeline and in the beginning of second timeline. Whereas Russia and United states fought against each other majorly in the third timeline.

\begin{table}[tp]
\caption{Rival Countries}
\resizebox{\textwidth}{!}{
\begin{tabular}{|l|l|c|l|l|c|}
\hline
\multicolumn{ 3}{|c|}{\textbf{[1500 – 1800]}} & \multicolumn{ 3}{c|}{\textbf{[1801 – 1945]}} \\  \hline
\multicolumn{ 1}{|c|}{\textbf{Node1}} & \multicolumn{ 1}{|c|}{\textbf{Node2}} & \multicolumn{ 1}{|c|}{\textbf{Edge Count}} &\multicolumn{1}{c|}{\textbf{Node1}} & \multicolumn{ 1}{|c|}{\textbf{Node2}} & \multicolumn{ 1}{|c|}{\textbf{Edge Count}} \\ \hline \hline
United Kingdom & France & \textbf{43} & United Kingdom & Germany & \textbf{22} \\ \hline
France & Spain & \textbf{40} & Russia & Germany & \textbf{20} \\ \hline
France & Holy Roman Empire & \textbf{36} & Russia & Italy & \textbf{21} \\ \hline
United Kingdom & Spain & \textbf{36} & United Kingdom & Italy & \textbf{21} \\ \hline
Ottoman Empire & Spain & \textbf{30} & Indonesia & Dutch East Indies & \textbf{21} \\ \hline \hline
\multicolumn{ 3}{|c|}{\textbf{[1946 – 2020]}} & \multicolumn{ 3}{c|}{\textbf{[1500 – 2020]}} \\  \hline
\multicolumn{ 1}{|c|}{\textbf{Node1}} & \multicolumn{ 1}{|c|}{\textbf{Node2}} & \multicolumn{ 1}{|c|}{\textbf{Edge Count}} &\multicolumn{1}{c|}{\textbf{Node1}} & \multicolumn{ 1}{|c|}{\textbf{Node2}} & \multicolumn{ 1}{|c|}{\textbf{Edge Count}} \\ \hline \hline
Russia & United States & \textbf{23} & United Kingdom & France & \textbf{61} \\ \hline
United States & China & \textbf{18} & United Kingdom & Spain & \textbf{50} \\ \hline
Al-Qaeda & United States & \textbf{17} & France & Spain & \textbf{46} \\ \hline
Thailand & Vietnam & \textbf{16} & France & Holy Roman Empire & \textbf{37} \\ \hline
China & Vietnam & \textbf{16} & Russia & France & \textbf{35} \\ \hline
\end{tabular}}
\label{tab_riv_t}
\end{table}

\subsection{Countries With highest number of wars fighting against common enemy}
Next we analyzed the countries who fought the most number of wars against the same side at the same time. If a pair of node is present in the rival country list and also in this list, this implies that the pair of nodes was fighting against each other and also against a common enemy at the same time. This would represent a clique like situation where many countries were fighting against each other. Whereas if two countries have fought most wars fighting against the same side and very few or no war against each other it give us a sense of friendship among the two. Table \ref{tab_fre_t} shows the list of such nodes.\\

\noindent \textbf{Timeline 1: Early Wars[1500-1800]:} Pair of France and Spain was present in the [1500-1800] section of Table~\ref{tab_riv_t}, and is also present in the same section for Table~\ref{tab_fre_t}. This follows the first scenario and shows that the countries were fighting against each other. This follows the reasoning that we gave while introducing the timeline that many European Countries were fighting against each other.\\

\noindent \textbf{Timeline 2: Pre-world wars [1801-1945]:} United Kingdom-France, Russia-United Kingdom, United Kingdom-United States all have very small number of wars against each others and large number of wars along the same side so this reflects friendship among these pair of nodes. This is also evident from the world war facts as United Kingdom, France and Russia were part of ally countries during second world wars. USA also joined ally countries during later part of second world war. India helped the United kingdom in the world wars so India-United Kingdom pair is also present in the list. We can see these stats clearly reflect the dominance of world wars in this timeline.\\

\noindent \textbf{Timeline 3: Post-World wars[1946-2020]:} All the nodes present in [1946-2020] section of Table~\ref{tab_fre_t} are powerful nations. We will see in the later section that these are the nodes which are fighting most against the different terrorist organizations as allies. \\

\begin{table}[tp]
\caption{War Against Common Enemy}
\resizebox{\textwidth}{!}{
\begin{tabular}{|l|l|c|l|l|c|}
\hline
\multicolumn{ 3}{|c|}{\textbf{[1500 – 1800]}} & \multicolumn{ 3}{c|}{\textbf{[1801 – 1945]}} \\ \hline
\multicolumn{1}{|c|}{\textbf{Node1}} & \multicolumn{1}{c|}{\textbf{Node2}} & \textbf{Number of Wars} & \multicolumn{1}{c|}{\textbf{Node1}} & \multicolumn{1}{c|}{\textbf{Node2}} & \textbf{Number of Wars} \\ \hline \hline
Spain & Holy Roman Empire & \textbf{44} & United Kingdom & France & \textbf{35} \\ \hline
Ottoman Empire & Turkey & \textbf{26} & United Kingdom & India & \textbf{29} \\ \hline
Portugal & State of Brazil & \textbf{23} & Russia & United Kindom & \textbf{28} \\ \hline
Spain & Papal States & \textbf{22} & United Kingdom & United States & \textbf{26} \\ \hline
France & Spain & \textbf{20} & Germany & Italy & \textbf{24} \\ \hline \hline
\multicolumn{ 3}{|c|}{\textbf{[1946 – 2020] }} & \multicolumn{ 3}{c|}{\textbf{[1500 – 2020] }} \\ \hline
\multicolumn{1}{|c|}{\textbf{Node1}} & \multicolumn{1}{c|}{\textbf{Node2}} & \textbf{Number of Wars} & \multicolumn{1}{c|}{\textbf{Node1}} & \multicolumn{1}{c|}{\textbf{Node2}} & \textbf{Number of Wars} \\ \hline \hline
United Kingdom & United States & \textbf{47} & United Kingdom & France & \textbf{79} \\ \hline
United States & Australia & \textbf{39} & United Kingdom & United States & \textbf{73} \\ \hline
United States France & France & \textbf{37} & United States France & France & \textbf{57} \\ \hline
United Kingdom & Australia & \textbf{33} & United Kingdom & Italy & \textbf{37} \\ \hline
United Kingdom & France & \textbf{32} & France & Spain & \textbf{35} \\ \hline
\end{tabular}}
\label{tab_fre_t}
\end{table}

\noindent \textbf{Overall Analysis [1500-2020]:}The reasoning given in the beginning of Section~4.4 fails here. Recall, we noted that if a pair of node is present in the rival country list and also in this list, this shows that the pair of nodes were fighting against each other and also against a common enemy at the same time. This represents a clique like situation where many countries are fighting against each other. Where as if two countries have fought most wars fighting against the same side and very few or no against each other it gives us a sense of friendship among the two. As the size of timeline is much larger, there is a possibility that two countries fought many wars against each other at some time and fought along the same side at a different time and this would not imply a clique like situation. Table \ref{tab_fre_t} shows the list of such nodes. Pairs at the top of the list are the pairs which fought together during the world wars and are also fighting against terrorism in the present era e.g United Kingdom-France, United Kingdom-United States, United states-France. 

\subsection{Inter-country Relations}
\begin{figure}[htbp]
\begin{subfigure}{.325\textwidth}
\includegraphics[width=\textwidth]{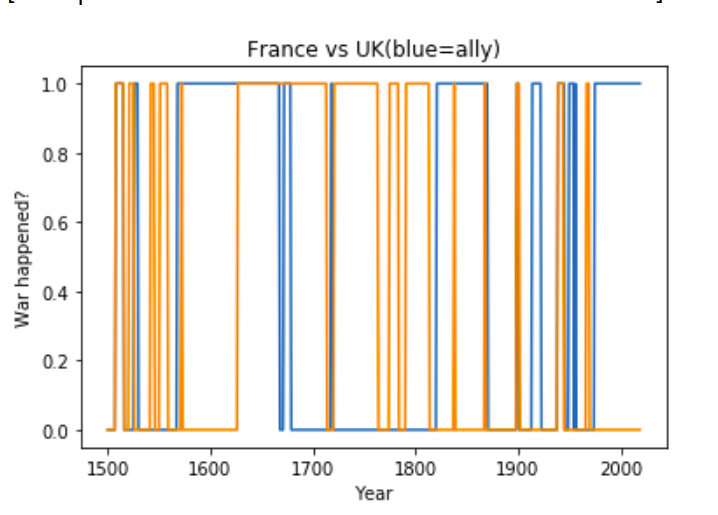}
\caption{UK and France}
\label{fig_uk_france}
\end{subfigure}%
\begin{subfigure}{.325\textwidth}
\includegraphics[width=\textwidth]{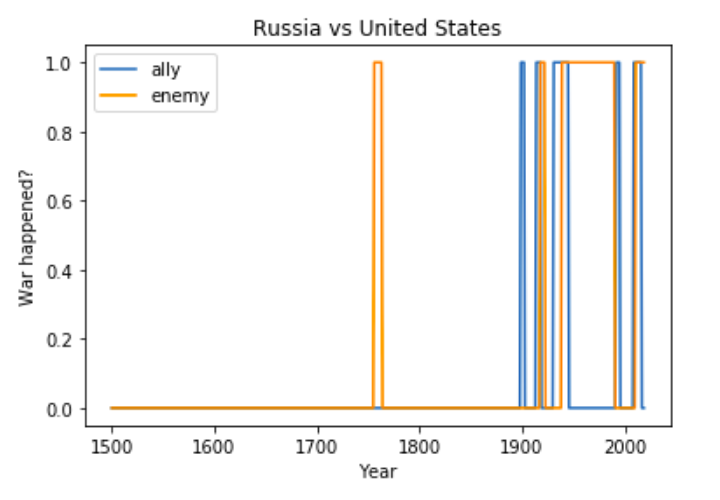}
\caption{USA and Russia}
\label{fig_us_russia}
\end{subfigure}
\begin{subfigure}{.325\textwidth}
\includegraphics[width=\textwidth]{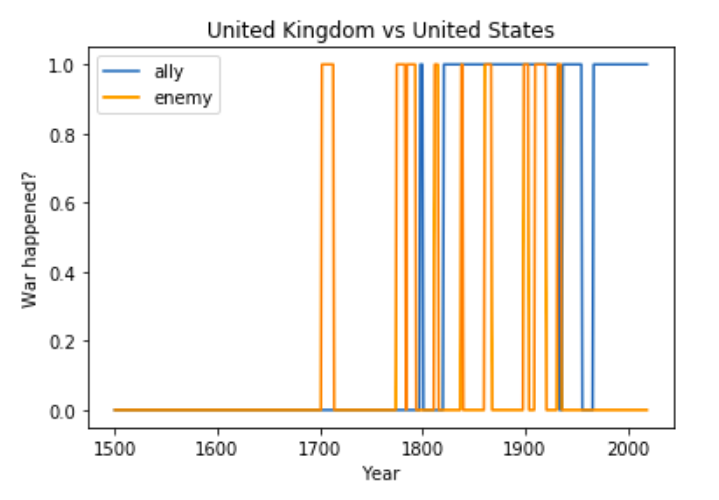}
\subcaption{UK and USA}
\label{fig_uk_us}
\end{subfigure}%
\label{fig_inter}

\begin{subfigure}{.325\textwidth}
\includegraphics[width=\textwidth]{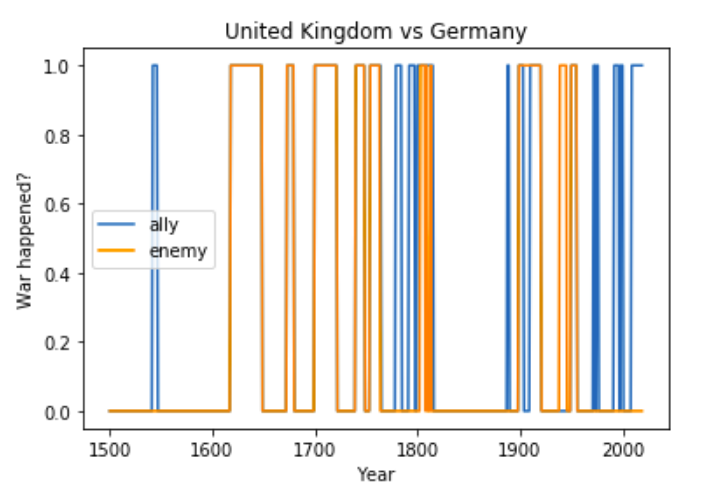}
\caption{UK and Germany}
\label{fig_uk_germany}
\end{subfigure}
\begin{subfigure}{.325\textwidth}
\includegraphics[width=\textwidth]{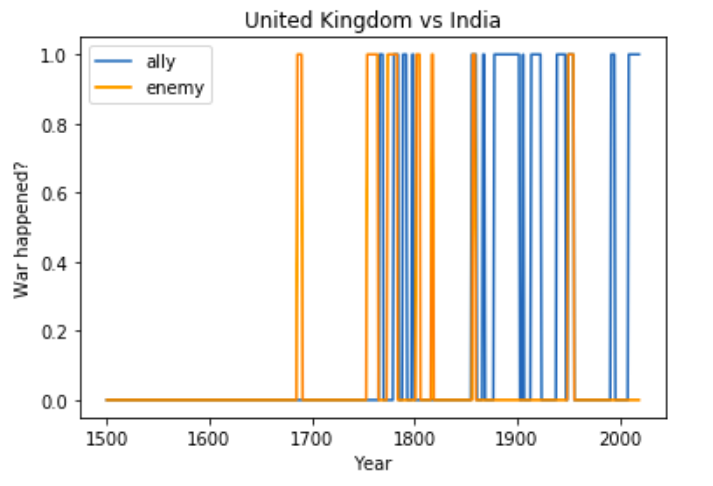}
\subcaption{UK and India}
\label{fig_uk_india}
\end{subfigure}
\begin{subfigure}{.325\textwidth}
\includegraphics[width=\textwidth]{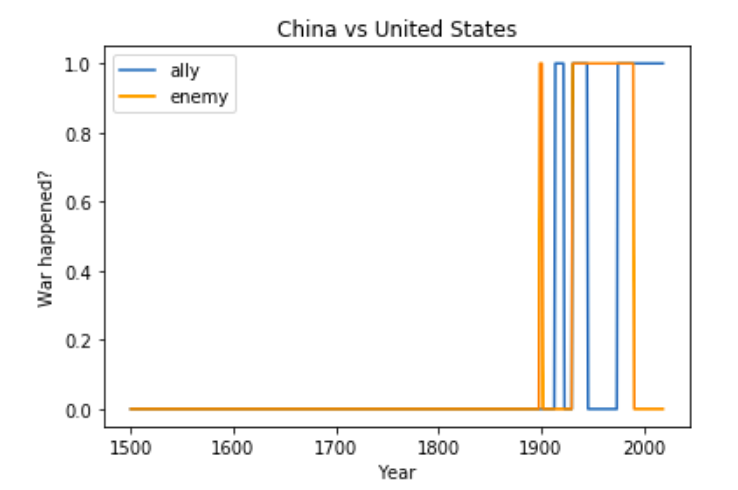}
\subcaption{USA and China}
\label{fig_us_china}
\end{subfigure}
\caption{Relations between two countries}
\label{fig_inter1}
\end{figure}

We studied the inter-country relations between different countries by analyzing the number of wars they fought against each other and number of wars they fought along the same side. To analyze this we plotted two lines for each pair of nodes. The blue line indicates if the two nodes fought a war along the same side in that year, where as the orange line indicates if two nodes fought against each other in that year. In the graph we plot both the lines and depending on the values of lines 4 scenarios are possible. Orange line is at 1 and blue line is at 0: Countries were fighting against each other in that particular year. Orange line is at 1 and Blue line is at 1: Countries were fighting against each other and also against some common opponent in that particular year. Orange line is at 0 and Blue line is at 1: Countries were fighting against some common opponent in that particular year. Orange line is at 0 and Blue line is at 0: Countries did not have any direct interaction with each other. Analysis of some inter-country relationship pairs are plotted in Figure ~\ref{fig_inter1}.\\

\noindent \textbf{United Kingdom and France:}
Figure~\ref{fig_uk_france} shows the plot of the relationship between United Kingdom and France based on the wars fought between them and against common enemies. Orange line has value 1 for almost all the time between 1500 CE and 1820 CE. This is because of the Anglo-French war which happened during that time. Blue line has value 1 for some time between 1550 CE to 1650 CE. This is because of the 80 years long war in which France and UK both fought against Spain from 1568 CE to 1648 CE. In 1830 France accepted Britain as ally so orange line doesn't touch y=1 value for long duration after that. Blue line touches y=1 constantly after 1970 as both developed nations were helping other countries in fighting civil wars and in mitigating terrorism.\\

\noindent \textbf{United Kingdom and Germany:} Figure~\ref{fig_uk_germany} plots the relationship between United Kingdom and Germany. Germany didn't fight many wars against other countries between 1500 CE to 1600 CE as a result except for one spike both orange and blue lines move along y=0 during that time. Between 1615 CE and 1775 CE blue and orange line move along each other at value 1 because during this timeline, European countries were fighting against each other for dominance. In later part of $18^{th}$ century, Britain and Prussia became allies, so only blue line touches y=1 in that area. Both blue and orange lines are at y=0 between 1800 CE and 1900 CE indicating not much direct interaction among two countries. Both the lines start to touch y=1 after 1900 as pre-world war activities were happening. Orange line has value 1 around 1915 and 1945 indicating the world war in which both the countries fought against each other. Blue line touches y=1 constantly after somewhere around 1980.\\

\noindent \textbf{United States and Russia:} Figure~\ref{fig_us_russia} plots the relationship between United States and Russia. There is no interaction between countries before 1900 CE except for a small period around 1770 CE which is because of involvement of Russia in United States' revolution. Blue lines touches y=1 around 1914 and 1945 as countries participated together in world wars. Orange line touches y=1 after 1946 up to 1991 because of the cold war. Blue line touches y=1 and remains there afterward as both nations were helping other countries in fighting civil wars and in mitigating terrorism.\\

\noindent \textbf{United Kingdom and India:} Figure~\ref{fig_uk_india} plots the relationship between United Kingdom and India. Their is no interaction between countries before 1700 CE. Orange line touches y=1 many times between 1700 CE and 1800 CE because of multiple Anglo-Indian wars. Orange line touches y=1 several times between 1800 CE and 1950 CE e.g. at 1847 CE because of the Revolutionary war of India. After 1800, Britain started gaining control over India and Indian troops fought along with them in certain wars. As a result we see blue line touching y=1 from around 1800 CE to 1950 CE, as India got freedom in 1947. Blue line again touches y=1 as both India and United Kingdom were fighting against terrorism.\\

\noindent \textbf{United States and China:} Figure~\ref{fig_us_china} plots the relationship between United States and China. Orange line touches y=1 for the first time at 1900 CE as in 1900 CE, China fought against USA during Boxer Rebellion. Blue line touches y=1 around the timeline of world wars as both USA and China fought on the same side. Orange line touches y=1 after the second world war, as USA and China supported opposing countries in different wars. Korean war is one such example in which USA supported South Korea and China supported North Korea. Blue line again touches y=1 after 1990.\\

\noindent \textbf{United States and United Kingdom:}Figure~\ref{fig_uk_us} plots the relationship between United States and United Kingdom. Choctaw (Native American) had conflict with British government in 1702 CE -1713 CE due to which Orange line touches y=1 at that time. Orange line touches y=1 again at around 1777 CE because of USA's revolutionary war. After the revolution, two countries became ally so we can see blue line touches y=1 again. Conflict among countries started again in 1812 CE (War of 1812) After that both countries remain ally for most of the time from 1812 CE to 2020 CE excluding some small duration. Some dispute happened between these. countries due to which orange line also touches y=1 for small intervals. Both USA and UK fought along the same sides in world war and remained allies afterward.\\%both the developed nations are helping other countries in fighting civil wars and in mitigating terrorism due to which blue line touches y=1 for most of the timeline.
% \subsubsection{China and Japan}
% \begin{figure}[h]
%   \centering
%   \includegraphics[width=\linewidth]{samples/japan_vs_china.png}
%   \caption{Relations between Japan and Chine}
%   \Description{Relations between Japan and China}
% \end{figure}

\subsection{Analysis of Countries' History}
\begin{figure}[htbp]
\begin{subfigure}{.325\textwidth}
\includegraphics[width=\textwidth]{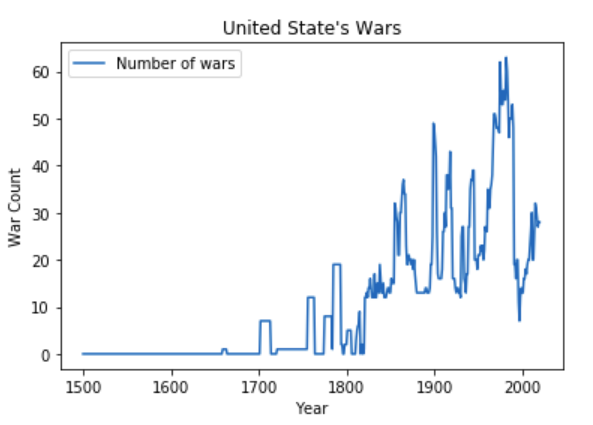}
\caption{United States}
\label{fig_us}
\end{subfigure}%
\begin{subfigure}{.325\textwidth}
\includegraphics[width=\textwidth]{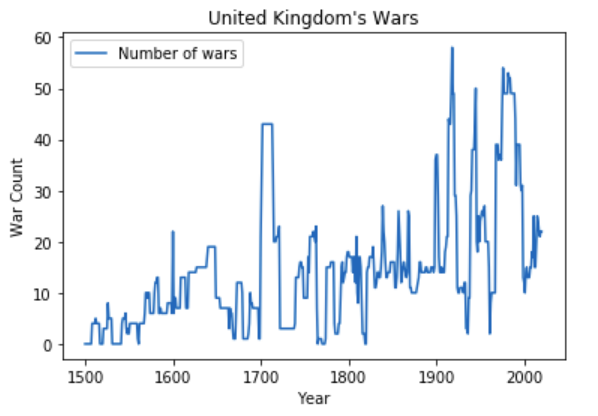}
\caption{United Kingdom}
\label{fig_uk}
\end{subfigure}
\begin{subfigure}{.325\textwidth}
\includegraphics[width=\textwidth]{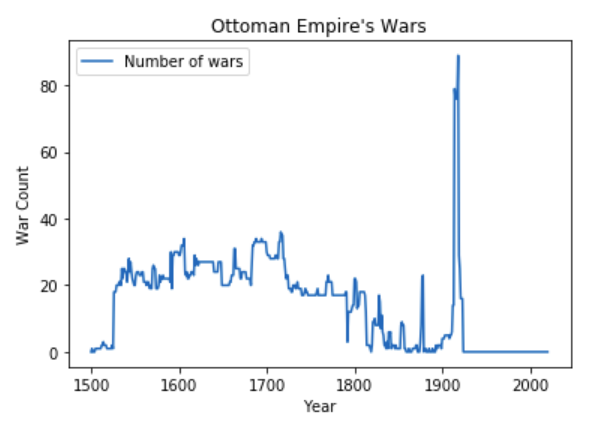}
\caption{Ottoman Empire}
\label{fig_ottoman}
\end{subfigure}
\label{fig_count_wars}
\begin{subfigure}{.325\textwidth}
\includegraphics[width=\textwidth]{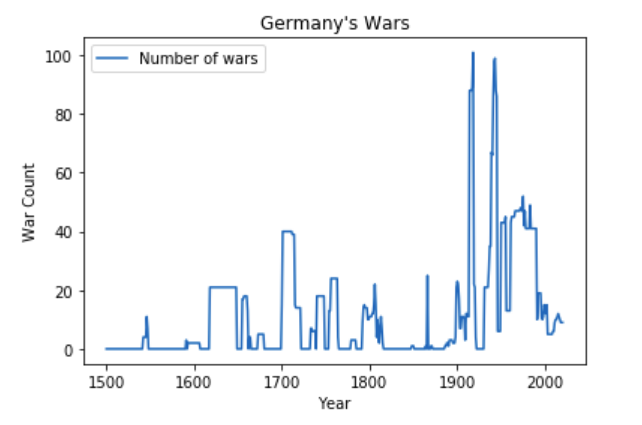}
\caption{Germany}
\label{fig_germany}
\end{subfigure}
\begin{subfigure}{.325\textwidth}
\includegraphics[width=\textwidth]{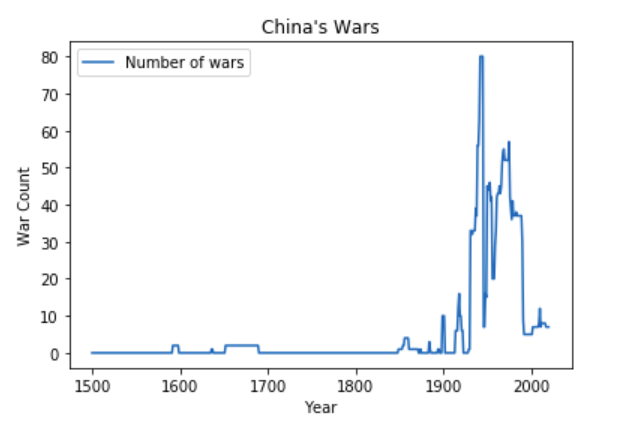}
\caption{China}
\label{fig_china}
\end{subfigure}
\begin{subfigure}{.325\textwidth}
\includegraphics[width=\textwidth]{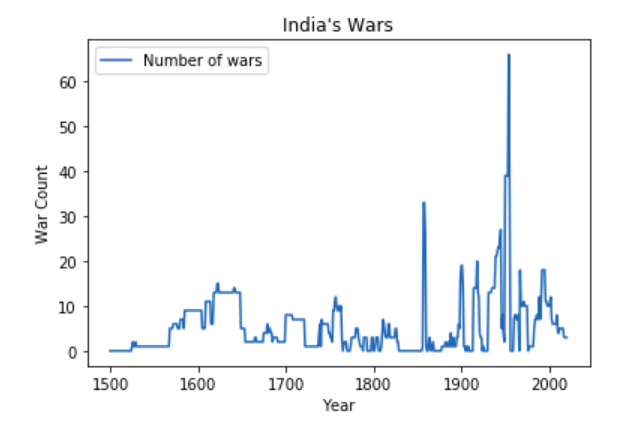}
\caption{India}
\label{fig_india}
\end{subfigure}%
\caption{Number of wars participated by countries/Empires over timeline}
\label{fig_count_wars1}
\end{figure}

% In the temporal network an edge only exist in the timeline of corresponding war. Some of the major events in the history of a country can be predicted by analysing its degree in the temporal network. 
This sections sheds light on the year-wise number of war-fronts in which some countries and empires were involved in. In these analysis, wars involving terrorist organizations are ignored to have a clear understanding of inter-country wars. The analysis for considered countries is plotted in Figure~\ref{fig_count_wars1}.\\

\noindent \textbf{United States:} Figure~\ref{fig_us} plots year-wise number of wars fought by United States. First significant increase in the curve is at 1702 CE - 1712 CE because of conflict between Native Americans and Britain. United States' revolution happened in 1776 CE so it hasn't participated in any war before that. After 1775 CE, curve starts increasing because of the Revolutionary War(1776). Curve shows significant increase around 1861 because of United States' Civil War. Two spikes in curve in the timeline of world wars. United States participated in various different wars during 1970 CE to 2000 CE e.g. Cold war, Gulf war, Tanker war, Tobruk wars etc. due to which curve attains its peak value around this time. United States is still sending its forces to help countries in need, therefore, curve remains at a significant value even after 2000 CE.\\

\noindent \textbf{Germany:} Figure~\ref{fig_germany} plots year-wise number of wars fought by Germany. First Considerable value is around 1610 CE to 1640 CE because of the 30 years war(1618-1658). Huge Spike in 1914 CE because of the first world war. Curve almost touches 0 after 1918 CE because of the restrictions imposed on Germany after the first world war. Another huge spike in 1945 CE because of the second world war. Number of wars significantly reduce after that.\\

\noindent \textbf{United Kingdom:} Figure~\ref{fig_uk} plots year-wise number of wars fought by United Kingdom. UK has been a strong country throughout the 520 years period in terms of fighting most number of wars. First significant increase in 1700 CE as United Kingdom participated in two big wars: Great Northern war,  and War of Spanish succession at the same time. The curve shows huge increase around the timeline of world wars reflecting the active participation of United Kingdom in it. United Kingdom participated in various different wars during 1970 CE to 2000 CE e.g. Bosnian war, Gulf war, Falklands wars, Cod wars etc.\\ %due to which curve attains high value even during this period.

\noindent \textbf{China:} Figure~\ref{fig_china} plots year-wise number of wars fought by China. The republic of china was formed in 1912 CE. Curve takes significant values after 1912. The curve takes value approximately 15 in 1914 CE reflecting China's participation in the first world war. Curve achieves its peak value around 1945-1955 period, as China fought against Japan in the second world war and also participated in battle of Chamdo, Korean war and the war against Taiwan. Curve has significant value throughout second half of 20th century reflecting its participation in Taiwan Crisis, Tibetan Uprising, Xinjiang conflict, Sino-Indian war, Vietnam war etc. As China is still sending its forces to help countries like Somalia as a result curve keeps a significant value even after 2000s.\\

\noindent \textbf{India:} Figure~\ref{fig_india} plots year-wise number of wars fought by India. The curve has lower values as compared to the above 4 plots. Small number of wars up to 1850 CE, because of wars between different empires (Maurya, Maratha, Sikkha, British Rule). First spike in 1857 CE because of Indian revolution movement during that time. Curve attains largest value around 1945 CE - 1955 CE period as India participated in the second world war to help Britain and got freedom in 1947 CE. India fought with Pakistan in 1947-48 CE for the state of Jammu and Kashmir. India also conducted military operations for annexation of Hyderabad and Junagadh. India had a conflict with Portugal for annexation of Dadra and Nagar Haveli in 1954. Curve doesn't vanish even though India isn't participating in war directly. It is because India is part of NATO which is helping different countries in fighting their wars.\\

\noindent \textbf{Ottoman Empire:} Figure~\ref{fig_ottoman} plots year-wise number of wars fought by Ottoman Empire. We can observe that the curve takes significant value after 1520 CE, as Ottoman Empire came in the total power after 1520 CE. This curve gains significantly larger value as compared to rest of 5 plots from 1520 CE to 1700 CE. Empire starts loosing power as the number of war starts reducing after 1710 CE. Curve has a huge spike in 1914 CE reflecting its participation in the first world war. The Empire was defeated in 1918 and officially ended in 1922 as a result the curve vanishes after that.

\subsection{Wars involving Terrorist Organisations} 
\begin{figure*}[htbp]

\begin{subfigure}{.325\textwidth}
\includegraphics[width=\textwidth]{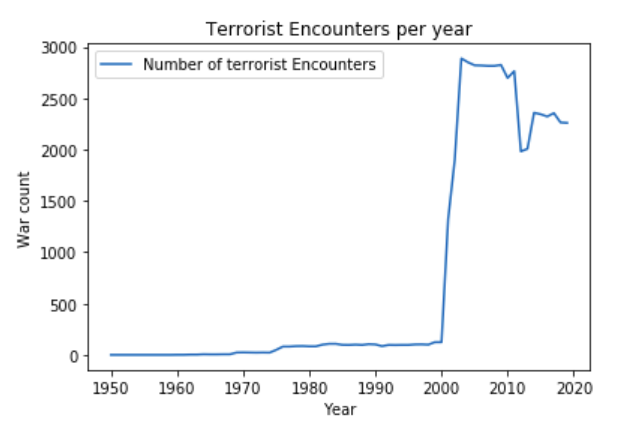}\\
\subcaption{Against all countries}
\label{fig_to}
\end{subfigure}%
\begin{subfigure}{.325\textwidth}
\includegraphics[width=\textwidth]{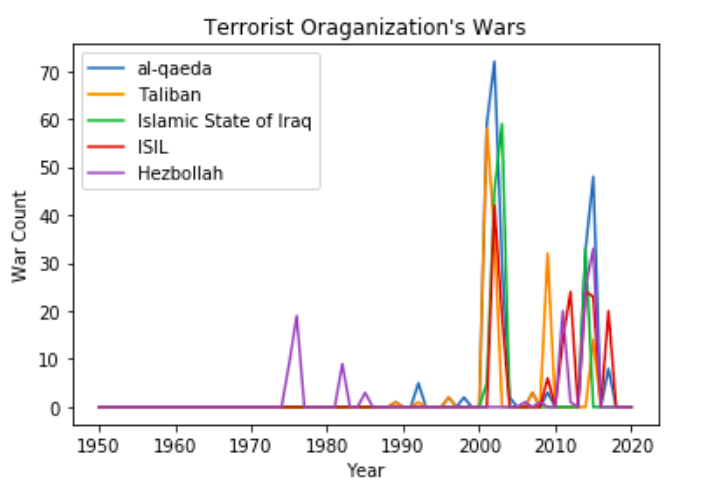}
\subcaption{For most involved terrorist organization}
\label{fig_top_to}
\end{subfigure}
\begin{subfigure}{.325\textwidth}
\includegraphics[width=\textwidth]{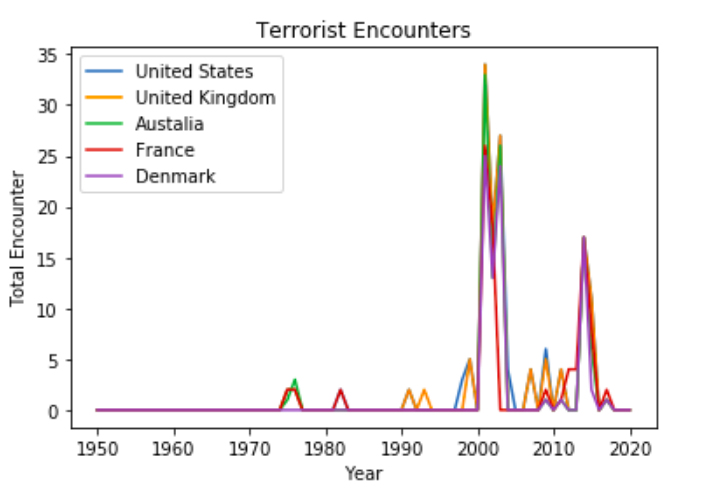}
\subcaption{Against most exposed countries}
\label{fig_top_count}
\end{subfigure}\\
\caption{Year-wise number of wars involving terrorist organizations}
\vspace{-2em}
\end{figure*}

In this section, we analyse the wars involving terrorist organization. These wars consider both: the attacks by terrorist organization on a country and attack by nations on the hubs of terrorist organizations. All the plots in this section is for timeline 1950 CE to 2020 CE because before 1950 CE, there is very limited number of such wars. Year-wise number of wars involving terrorist organization is plotted in Figure~\ref{fig_to}. It is observed that the total number of wars involving terrorist organization are close to zero in the begin. The curve increases slowly till 2000. Curve increases sharply and attains its maximum value in 2001. The number of terrorist encounters have decreased since then. Still the count is high.

The list of the top 5 terrorist organisations which have been involved in most wars. It reports that Al-Qaeda has been the most active terrorist organization. Table~\ref{tab_terror} also lists the top 5 countries with maximum engagement with various terrorist organisations in wars along with the number of wars. The timeline analysis for these top five entities is given in Figure~\ref{fig_top_to}. It becomes clear from these plots, when these organization became active and when they are at peak for getting involved in wars.

\begin{table}[h]
\vspace{-2em}
\caption{Wars involving Terrorist Organisations}
\resizebox{\textwidth}{!}{
\begin{tabular}{|l|c|l|c|}
\hline
\multicolumn{ 2}{|l|}{\textbf{Most Active Terrorist Organizations}} & \multicolumn{ 2}{l|}{\textbf{Countries with Most Terrorist Encounters}} \\ \hline
\textbf{Terrorist Organisation } & \textbf{Total Number of Wars} & \textbf{Country} & \textbf{Number of terrorist Encounters} \\ \hline \hline
Al-Qaeda & \textbf{422} & United States & \textbf{212} \\ \hline
Taliban & \textbf{235} & United Kingdom & \textbf{201} \\ \hline
Islamic States of Iraq & \textbf{208} & Australia & \textbf{168} \\ \hline
ISIL & \textbf{199} & France & \textbf{149} \\ \hline
Hezbollah & \textbf{148} & Denmark & \textbf{135} \\ \hline
\end{tabular}}
\label{tab_terror}
\vspace{-1em}
\end{table}

 Table~\ref{tab_top_pair} contains the number of engagement a country had in wars with each specific terrorist organizations. It turns out that Al-qaeda and United States are the top engaging nodes. Al-qaeda is the terrorist organisation in all top five node pairs that exhibits the reach and engagement of this organization across several continent. Al-Qaeda has been the most active terrorist organization. Further, it also helps us to analyse the names of the countries which are fighting most in trying to mitigate the terrorism. It reports that United States has participated most in the wars against terrorist organizations. It becomes clear from these plots, when these countries were first exposed against terrorism and when these get most engaged in wars against terrorist organization. Finally, we analyse the number of engagement a country had in wars with each specific terrorist organizations. It turns out that Al-qaeda and United States are the top engaging nodes. Al-qaeda is the terrorist organisation in all top five node pairs that exhibits the reach and engagement of this organization across several continent.
\begin{center}

\begin{table}[h]
\vspace{-3em}
  \caption{Most wars for Country-Terrorist Org. pairs:}\centering
\resizebox{.7\textwidth}{!}{\centering
\begin{tabular}{|c|c|c|}
\hline
\textbf{Country} & \textbf{Terrorist Organisation} & \textbf{Number of encounters} \\
\hline
United States & Al-Qaeda & 18\\ \hline
United Kingdom & Al-Qaeda & 13\\\hline
Australia  & Al-Qaeda & 12\\\hline
Netherland & Al-Qaeda & 11\\\hline
Denmark & Al-Qaeda & 10\\
\hline
\end{tabular}}
\label{tab_top_pair}
\vspace{-4em}
\end{table}
\end{center}

\subsection{Economic Effects of War}
\begin{figure}[tbp]
\begin{subfigure}{.325\textwidth}
\centerline{\includegraphics[scale=0.4]{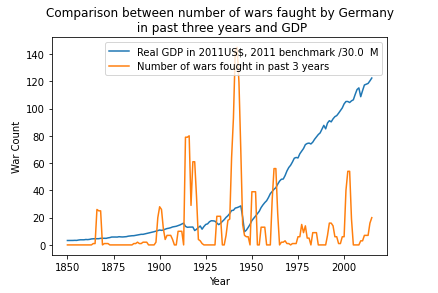}}
\caption{Germany}
\label{fig}
\end{subfigure}
\begin{subfigure}{.325\textwidth}
\centerline{\includegraphics[scale=0.4]{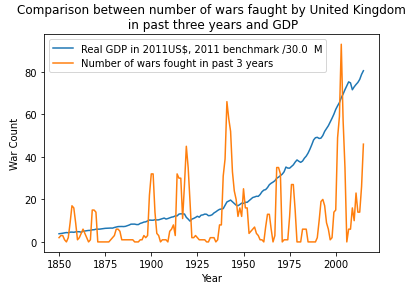}}
\caption{United Kingdom}
\label{fig}
\end{subfigure}
\begin{subfigure}{.325\textwidth}
\centerline{\includegraphics[scale=0.4]{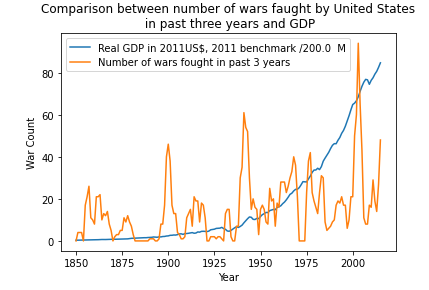}}
\caption{USA}
\label{fig}
\end{subfigure}
\caption{Economic Effects of War}
\label{fig:sfig}
\vspace{-2em}
\end{figure}

%We also analyse how wars impacted the economic conditions of fighting countries. 
We have plotted the number of wars fought by each country and their GDP value and we can clearly see the correlation among the two. We have plotted the two stats represented by blue and orange line across the timeline (X axis). The blue line represents the GDP of the country in that particular year. We have scaled it down by dividing with a constant to show correlation properly. The Orange line represents the number of wars fought by the country in past 3 years. We have plotted the cumulative sum of wars for past 3 years instead of the wars in that particular year since wars impact the GDP of upcoming years as well. We observe that the spikes in the orange lines are always followed by a dip in the blue curve for all countries. We have attached few of those plots in the Figure~\ref{fig:sfig}. The dips are deeper for few countries and shallower for the others. Note that all the countries have a dip around 1945-50 because of the second world war but the size of dip (relative to their GDP) for United States and United Kingdom is smaller in comparison to that of Germany and Japan. It seems that the countries which won the wars tend to have smaller dips. The impact of outcome of the wars require further investigation.

\section{Conclusion}
Temporal analysis of wars fought after 1500 CE is done by curating data available on Wikipedia. The analysis results comply with the historical events. The powerful and dominating nations are identified during different timelines. Rivalries and allies based on wars explain the inter-country relationships during different timelines. Most active terrorist organizations and most active countries against terrorism were also identified using this analysis. The data did not include the damage associated with wars, due to which all wars were treated equally. Analysis incorporating damage associated with wars by assigning weights on the edges in temporal graph and considering the decision of war will give better insights. Impact on economy has been analysed in terms of change in GDP.  After 1500, european countries began to move from feudalism to capitalist societies, and domestic revolutions played a crucial role in the future development of the countries. Taking into account the industrial revolution and the impact of domestic revolutionary movements on GDP is another future direction.

\bibliographystyle{splncs04}
\bibliography{acmart}

\begin{thebibliography}{10}
\providecommand{\url}[1]{\texttt{#1}}
\providecommand{\urlprefix}{URL }
\providecommand{\doi}[1]{https://doi.org/#1}

\bibitem{wiki}
https://en.m.wikipedia.org/wiki/category:lists\_of\_wars\_by\_country (2020)

\bibitem{baffigi2011italian}
Baffigi, A.: Italian national accounts, 1861-2011. Bank of Italy Economic
  History Working Paper (18) (2011)

\bibitem{RePEc:hit:hitcei:2018-13}
Bassino, J.P., Broadberry, S., Fukao, K., Gupta, B., Takashima, M.: {Japan and
  the Great Divergence, 730-1874}. CEI Working Paper Series 2018-13, Center for
  Economic Institutions, Institute of Economic Research, Hitotsubashi
  University (Dec 2018),
  \url{https://ideas.repec.org/p/hit/hitcei/2018-13.html}

\bibitem{https://doi.org/10.1111/1468-0289.12032}
Bolt, J., van Zanden, J.L.: The maddison project: collaborative research on
  historical national accounts. The Economic History Review  \textbf{67}(3),
  627--651 (2014)

\bibitem{BROADBERRY201558}
Broadberry, S., Custodis, J., Gupta, B.: India and the great divergence: An
  anglo-indian comparison of gdp per capita, 1600–1871. Explorations in
  Economic History  \textbf{55},  58--75 (2015).
  \doi{https://doi.org/10.1016/j.eeh.2014.04.003},
  \url{https://www.sciencedirect.com/science/article/pii/S0014498314000187}

\bibitem{broadberry2010british}
Broadberry, S.N., Campbell, B.M., Klein, A., Overton, M., Leeuwen, B.v.:
  British economic growth: 1270-1870  (2010)

\bibitem{RePEc:cup:cbooks:9780521817912}
Carter, S.B., Gartner, S.S., Haines, M., Olmstead, A., Sutch, R., Wright, G.
  (eds.): The Historical Statistics of the United States 5 Volume Hardback Set.
  Cambridge University Press (2006),
  \url{https://EconPapers.repec.org/RePEc:cup:cbooks:9780521817912}

\bibitem{10.2307/27744128}
de~la Escosura, L.P.: Lost decades? economic performance in post-independence
  latin america. Journal of Latin American Studies  \textbf{41}(2),  279--307
  (2009), \url{http://www.jstor.org/stable/27744128}

\bibitem{Glick:2010}
Glick, R., Taylor, A.M.: Collateral damage: Trade disruption and the economic
  impact of war. The Review of Economics and Statistics  \textbf{92}(1),
  102--127 (2010)

\bibitem{Hafner:2009}
Hafner-Burton, E.M., Kahler, M., Montgomery, A.H.: Network analysis for
  international relations. International Organization  \textbf{63}(3),
  559–592 (2009)

\bibitem{Homero:2018}
Homero Roman~Roman, Colin P.~Gaffney, L.F.V.: Temporal analysis of intenational
  relations networks (2018),
  \url{http://snap.stanford.edu/class/cs224w-2018/reports/CS224W-2018-99.pdf}

\bibitem{Henk:1988}
Houweling, H., Siccama, J.G.: Power transitions as a cause of war. Journal of
  Conflict Resolution  \textbf{32}(1),  87--102 (1988)

\bibitem{Jackson:2015}
Jackson, M.O., Nei, S.: Networks of military alliances, wars, and international
  trade. Proc. of the National Academy of Sciences  \textbf{112}(50),
  15277--15284 (2015)

\bibitem{Lagazio:2004}
Lagazio, M., Russett, B.: A neural network analysis of militarized disputes,
  1885-1992: Temporal stability and causal complexity (2004)

\bibitem{malanima2011long}
Malanima, P.: The long decline of a leading economy: Gdp in central and
  northern italy, 1300--1913. European Review of Economic History
  \textbf{15}(2),  169--219 (2011)

\bibitem{David:1999}
Mason, T.D., Joseph P~Weingarten, J., Fett, P.J.: Win, lose, or draw:
  Predicting the outcome of civil wars. Political Research Quarterly
  \textbf{52}(2),  239--268 (1999)

\bibitem{Mcmahan:2005}
McMahan, J.: Just cause for war. Ethics \& International Affairs
  \textbf{19}(3),  1–21 (2005)

\bibitem{Murthy:2006}
Murthy, R.S., Lakshminarayana, R.: Mental health consequences of war: a brief
  review of research findings. World psychiatry  \textbf{5}(1), ~25 (2006)

\bibitem{Organski:1981}
Organski, A., Kugler, J.: The War Ledger (06 1981)

\bibitem{pfister2011economic}
Pfister, U.: Economic growth in germany, 1500--1850. In: conference on
  quantifying long run economic development, Venice. pp. 22--24 (2011)

\bibitem{Karl:2004}
Karl R.~de Rouen, J., Sobek, D.: The dynamics of civil war duration and
  outcome. Journal of Peace Research  \textbf{41}(3),  303--320 (2004)

\bibitem{Sarkees:2010}
Sarkees, M.R., Wayman, F.: Resort to war: 1816 - 2007 (2010),
  \url{https://correlatesofwar.org/data-sets/COW-war}

\bibitem{https://doi.org/10.1111/1475-4932.12250}
Tang, J.P.: Review: Regional inequality and industrial structure in japan:
  1874-2008, by kyoji fukao, jean-pascal bassino, tatsuji makino, ralph
  papryzycki, tokihiko settsu, masanori takashima and joji tokui (maruzen
  publishing company, ltd, tokyo, 2015), pp. 350. Economic Record
  \textbf{92}(296),  141--143 (2016).
  \doi{https://doi.org/10.1111/1475-4932.12250},
  \url{https://onlinelibrary.wiley.com/doi/abs/10.1111/1475-4932.12250}

\end{thebibliography}
\end{document}